# Bright oxygen- and vacancy-derived spin-singlet diamond color centers with metastable spin triplets: $OV^{2+}$ and $VOV^{2+}$

John Mark P. Martirez*
Applied Materials and Sustainability Sciences, Princeton Plasma Physics Laboratory, Princeton, New Jersey, USA
*martirez@pppl.gov, martirez@princeton.edu

**Abstract**

The ST1 diamond color center was experimentally demonstrated to involve a substitutional oxygen atom ($O_C$) and carbon vacancy ($V_C$), has a spin singlet ground-state, and a metastable electron spin ancilla: a triplet. ST1's structure was left unsolved for more than a decade. With embedded multiconfigurational quantum mechanical theory, we investigate $O_C$-$V_C$-derived diamond defects, specifically both 0 and +2-charged coupled $O_CV_C$, and $O_C$ surrounded by $V_C$s along the [110] axis ($V_CO_CV_C$). We found both $O_CV_C^{2+}$ ($C_{3v}$) and $V_CO_CV_C^{2+}$ ($C_{2v}$) to have a spin-singlet ground state ($1^1A_1$) and metastable spin triplets. We demonstrate ST1 to be $V_CO_CV_C^{2+}$. The calculated vertical excitation energies of $V_CO_CV_C^{2+}$'s first ($1^1B_2$) and second ($2^1A_1$) bright spin-singlet excited states closely match ST1's experimental zero phonon line (2.2-2.3 eV). $O_CV_C^{2+}$ ($^1E$) absorbs much higher (2.8 eV). The two O lone pairs favor $V_CO_CV_C$ over $O_CV_C$, in a similar manner as the single N lone pair favors formation of $N_CV_C$ centers.

## Main

Single-photon emitters based on specialized defects in diamond (color centers) are at the forefront of architectures for building quantum devices, e.g., quantum computers and quantum sensors.[1-3] Such a defect or their ensemble constitutes a quantum bit where electron and nuclear spins, both of which can be subject to a coherent superposition of state, form the basis for information writing, processing, and readout.[1-3]

Oxygen-derived defects are currently receiving attention as promising diamond color centers to be used as qubits for quantum information applications, notably the ST1 color center.[4-7] The discovery of ST1 was first documented more than a decade ago (2013) from reactive ion etching with O plasma of single-crystal diamond vertical nanowires.[4] It was then the second known solid-state single spin defect with detectable (nuclear) spin coherence at room temperature, in addition to the well-characterized negatively charged NV defect: a substitutional nitrogen ($N_C$) adjacent to a carbon vacancy ($V_C$) or $N_CV_C^-$.[4] ST1 possesses longer spin resonance lifetime than $N_CV_C^-$, and has a spin singlet ground-state and a metastable electron spin ancilla: a triplet,[4] which is opposite of $N_CV_C^-$. Intrinsic hyperfine splitting is absent in ST1, indicating that the defect's constituent atoms have no nuclear spin, which is true for abundant isotopes $^{16}O$ and $^{12}C$, but not, e.g., $^{14}N$. Conversely, the non-magnetic ground state of ST1 defect makes it a desirable quantum bus (long-live storage) that registers nuclear spin upon introduction of $^{13}C$ (no nuclear spin decoherence resulting from net electron spins).[5] High-contrast readout is then achieved by optical excitation that populates the electron spin triplet.[4] Experimentally demonstrated to be associated with a substitutional oxygen atom ($O_C$), ST1 likely also involves $V_C$ and has a $C_{2v}$ or lower ($C_s$) symmetry with a [110] defect axis.[4-7] Multiple studies on ST1 (and related oxygen-derived defects) have been published since its discovery aiming to further characterize and synthesize it more consistently and robustly, as well as determine its exact composition, structure, and charge state.[5-9] However, after more than a decade since the defect's first discovery, its structure remains unsolved. Using accurate quantum mechanical simulations, here we present the identity of ST1 backed by multiple pieces of evidence based on (1) composition, (2) structural symmetry, (3) optical absorption profile: excitation energies, brightness, and emission lifetime, (4) magnetic properties: spin multiplicity and zero-field splitting, and (5) $O_C$-$V_C$ coupling thermodynamics, which were compared (many quantitatively) to experimental literature.[4-7] We use our rigorously benchmarked modelling approach,[10] namely, the capped density functional embedding theory (capped-DFET)[11,12] with embedded correlated wavefunction theory[13] to quantum mechanically model isolated defects with accurate excited state theories. Here we focus on two possible $O_C$-$V_C$ coupled configurations with ratios of 1:1 and 1:2, and defect charges 0 and +2 that yield even number of electrons, given ST1's spin singlet and triplet ground and excited states, respectively.

## Results

*Atomic structures of $O_CV_C$ and $V_CO_CV_C$ defects*. $O_CV_C$ is structurally similar to the $N_CV_C$ defect, where the substitutional $O_C$ sits at a C lattice site adjacent to a $V_C$. This configuration forces $O_C$ to form three bonds with C: two covalent and one coordinate covalent where O is the electron pair donor. Just like $N_CV_C$, $O_CV_C$, may adopt $C_{3v}$ symmetry with a three-fold axis of rotation ($C_3$) along the $O_C$–$V_C$ path - **Fig. 1A**. $N_CV_C^-$ and $O_CV_C^0$ are also iso-electronic and thus possess similar electronic structures (vide infra). $V_CO_CV_C$ on the other hand involves two $V_C$s and an O replaces the C atom shared by the $V_C$s. The O atom sits at a two-fold binding site and forms the ideal two covalent bonds (**Fig. 1B**). $V_CO_CV_C$ may adopt $C_{2v}$ symmetry as its highest possible symmetry, with a notable two-fold ($C_2$) axis of rotation ($z$) along [001] and centered at O, bisecting the C-O-C bond angle (**Fig. 1B**). The $V_C$–$V_C$ path is along an orthogonal axis ($y$), i.e., [110]. $C_{2v}$ has two reflection symmetries ($\sigma_v$): the plane either containing ($xz$) or bisecting ($yz$) C-O-C (**Fig. 1B**). $V_CO_CV_C$ is distinct from the split vacancy structure of the SiV defect: an interstitial Si ($Si_i$) surrounded by two $V_C$s ($V_CSi_iV_C$) that has $D_{3d}$ symmetry and a [111] defect axis.[14]

We model $O_CV_C$ and $V_CO_CV_C$ using bulk $C_{126}O$ and $C_{247}O$ periodic supercells, respectively, optimized within density functional theory (DFT) and the $r^2$-SCAN-L exchange correlation (XC) functional[15] (**Methods**). From these bulk structures we carved out $C_{15}O$ and $C_{26}O$ clusters, respectively, to which we add atoms at the periphery of the cluster to coordinatively saturate the C atoms at the edges. F and O were added for terminal (monovalent) and bridging (divalent) capping, respectively, to yield $C_{15}OF_{12}O_{12}$ and $C_{26}OF_{18}O_{15}$ clusters. **Fig. 2** shows these models which were chosen such that the clusters retain the symmetry of the defects in the crystal (**Fig. 1A,B**).

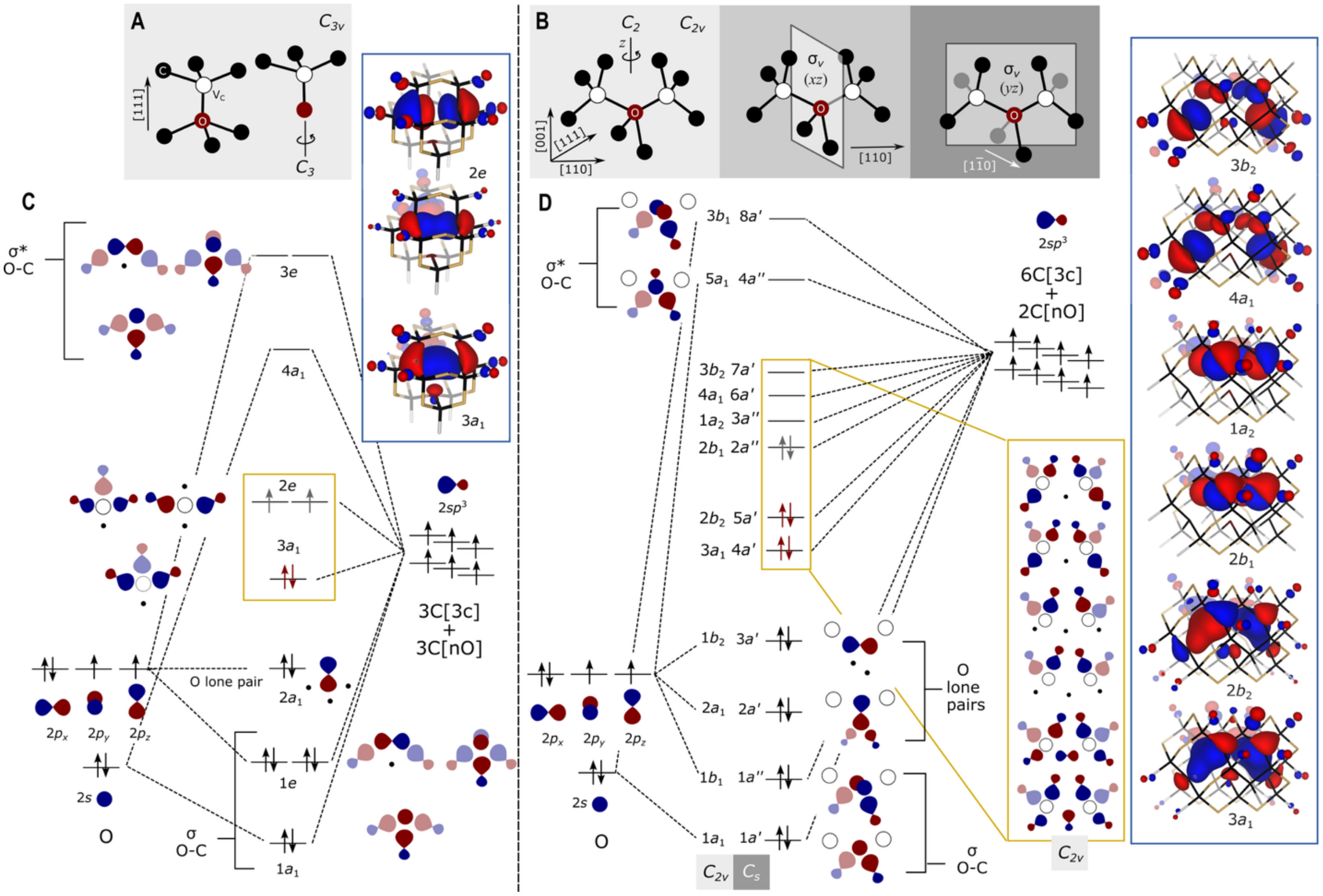

**Fig. 1. Defect-centered structural symmetry and molecular orbital (MO) diagrams.** (**A**,**B**) Highest possible symmetry of the defect sites are illustrated: $C_{3v}$ and $C_{2v}$ for $O_CV_C$ and $V_CO_CV_C$, respectively. The rotational axes are shown in both, and in $V_CO_CV_C$ the reflection planes ((110) and (1$\bar{1}$0) planes) are illustrated. MO diagrams for $O_CV_C$ (**C**) and $V_CO_CV_C$ (**D**) defects describing the orbital mixing between the substituent O and the surrounding C atoms (C[3c]s and C[nO]s, **Fig. 2**). Orbital labels are shown consistent with the defects' point group symmetries. The valence orbitals (gold boxes) give rise to the multiconfigurational nature of the defects (so-called active space) and are primarily involved in the lowest energy excitations. For $O_CV_C^0$ and $V_CO_CV_C^0$, these orbitals respectively contain four and six electrons, which reduces to two and four electrons for the +2-charged defects. Insets (blue boxes) show the calculated state-averaged active-space canonical orbitals from emb-CASSCF simulations for the spin-singlet +2 charged defects. They naturally exhibit symmetries expected for their point group, without explicit symmetrization.

$O_CV_C$ and $V_CO_CV_C$ feature three and six three-fold coordinated C atoms (C[3c]). We distinguish two type of C[3c]: type I has two C[3c] neighbors, whereas type II has three C[3c] nearest neighbors. Type II is only found in $V_CO_CV_C$ (**Fig. 2**). $O_C$ has respectively three and two nearest C atoms (C[nO]) in $O_CV_C$ and $V_CO_CV_C$ (**Fig. 2**). We quantum mechanically included the effect of the periodic diamond crystal upon the capped defect clusters via an optimized effective potential within DFT with HSE06 XC functional[16] following the capped-DFET formalism[11,12] (**Methods**). Using the clusters in **Fig. 2**, embedded in capped-DFET-derived potential, we performed multiconfigurational wavefunction theory simulations of the defects, specifically, the complete active space self-consistent (CASSCF) method[17] in conjunction with the *n*-electron valence second order perturbation theory (NEVPT2),[18] which would be impossible to perform on large extended models. Hereafter, we will refer to these simulations as emb-CASSCF and emb-NEVPT2.

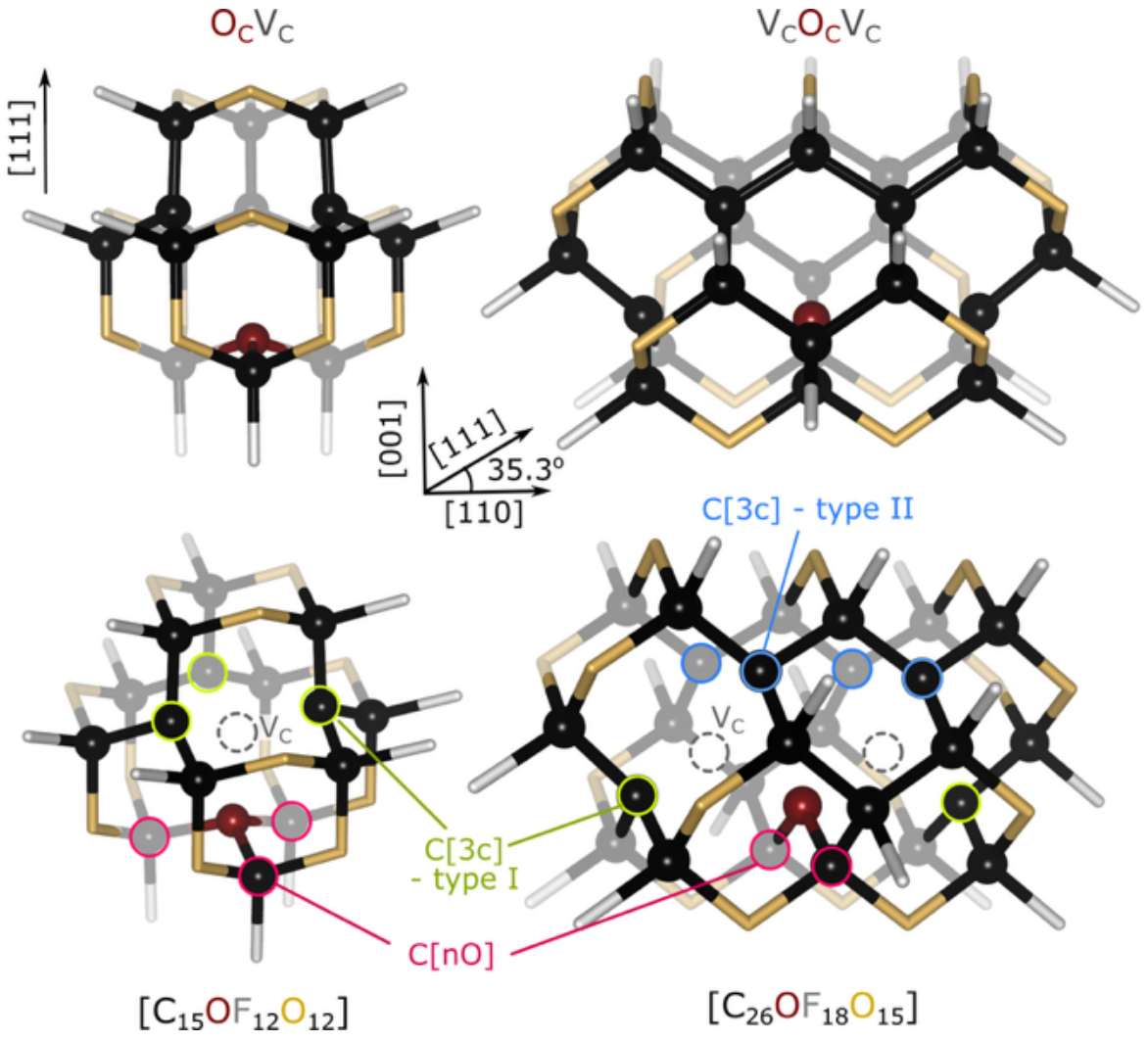

**Fig. 2. Capped cluster models**. The $C_{15}O$ ($O_CV_C$) and $C_{26}O$ ($V_CO_CV_C$) clusters were carved from a $C_{126}O$ and $C_{247}O$ periodic diamond structure, respectively. The edges of the clusters were capped with either F, if capping a terminal C atom, or O, if capping simultaneously two C atoms (bridging). The capping elements were placed where lattice C should be in the crystal and no further relaxation performed. The bottom structures are angled views of the structures shown on top. C - black, O defect - dark red, O cap - yellow, and F cap - white. The vectors shown are in the basis of the cubic diamond lattice vectors.

*Optical and magnetic properties of $O_CV_C$ and $V_CO_CV_C$.* **Fig. 1C** shows the valence orbital mixing in $O_CV_C$ (within the molecular orbital (MO) picture) between $O_C$, three C[3c]s, and three C[nO]s (**Fig. 2**). In **Fig. 1C**, the bonding σ orbitals are labelled as $1a_1$ and $1e$, and the anti-bonding $\sigma^*$ are $4a_1$ and $3e$ - following the Mulliken orbital notation based on $C_{3v}$ symmetry. $O_C$ also has one free lone pair labelled as $2a_1$. The three C[3c]s that form the basis of $V_C$ engender the non-bonding orbitals: $3a_1$ (all non-bonding C $sp^3$ orbitals are in-phase) and a doubly degenerate $2e$ (non-bonding C $sp^3$ orbitals are out-of-phase). The (primary) ground-state electron configurations of the +2 and neutral $O_CV_C$ are $3a_1^22e^0$ ($s = 0$, spin singlet) and $3a_1^22e^2$ ($s = 1$, spin triplet), respectively. The calculated set of active-space canonical orbitals from spin-singlet-three-state averaged emb-CASSCF calculation for the $O_CV_C^{2+}$ cluster (**Fig. 2**) illustrate the $3a_12e$ symmetries (**Fig. 1C**, inset).

**Fig. 1D** shows the corresponding MO diagram for $V_CO_CV_C$, which shows valence orbital mixing between $O_C$, six C[3c]s, and two C[nO]. Each of the $V_C$s in $V_CO_CV_C$ have the same set three non-bonding orbitals that may linearly combine to yield six non-degenerate non-bonding orbitals of symmetries: $3a_1$ and $2b_2$ (linear combinations of two $3a_1$ orbitals in $O_CV_C$); and $2b_1$, $1a_2$, $4a_1$, and $3b_2$ (linear combinations of four 2$e$ orbitals in $O_CV_C$). Additionally, $O_C$ has two lone pair (2$a_1$1$b_2$) and yields two sets of O-C σ,σ* (1$a_1$1$b_1$5$a_1$3$b_1$) orbitals. **Fig. 1D** also show the corresponding orbital labels for a $C_s$ symmetry (a subgroup of $C_{2v}$) with a $(1\bar{1}0)$ reflection plane ($\sigma_v(1\bar{1}0)$): either $a$' (symmetric) or $a$" (antisymmetric). The ground-state (primary) electron configurations of spin-singlet $V_CO_CV_C^{2+/0}$ are (+2) $3a_1^2 2b_2^2 2b_1^0 1a_2^0 4a_1^0$ and (0) $3a_1^2 2b_2^2 2b_1^2 1a_2^0 4a_1^0$. The active-space canonical orbitals from spin-singlet eight-state averaged emb-CASSCF calculation for the $V_CO_CV_C^{2+}$ cluster (**Fig. 2**) naturally exhibit these symmetries (**Fig. 1D**, inset).

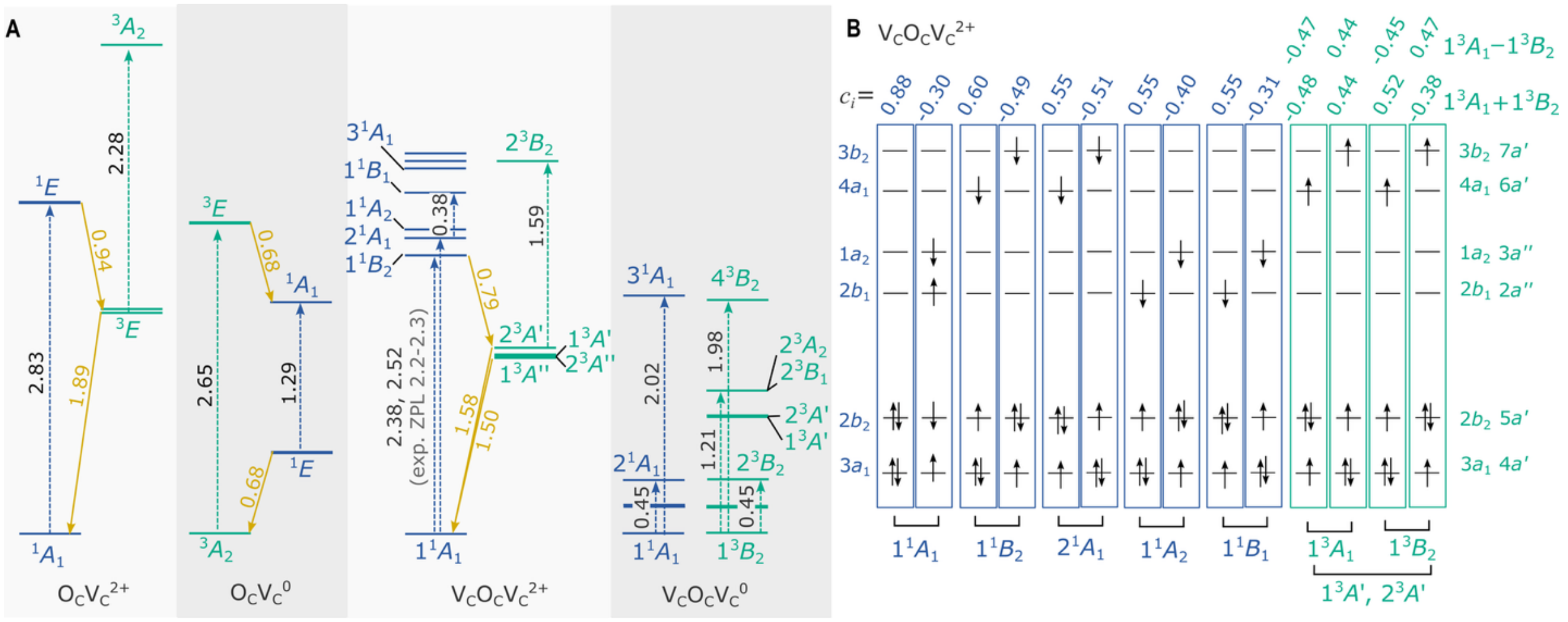

**Fig. 3. Jablonski (energy) diagrams and electron configurations. A**, Emb-NEVPT2 vertical excitation energies (in eV, to scale) for the oxygen-vacancy-derived defects (as labelled) showing the spin-singlet (blue) and -triplet (green) spectra. All calculated states are shown but only some are labelled. Some spin-allowed excitations are marked (blue and green up arrows). Spin-forbidden transitions (intersystem crossing) are marked with gold arrows. **B**, ground- and select excited-state primary electronic configurations for spin singlet and triplet $V_CO_CV_C^{2+}$ in the canonical orbital bases (Fig. 1D inset, Fig. S1). The configuration interaction coefficients ($c_i$) are given for each configuration $i$. For each state, $\sum_i |c_i|^2 = 1$ when summed over all configurations. The orbitals are labelled based on $C_{2v}$ ($s$ = 0,1) and $C_s$ ($s$ = 1) point group symmetries (Fig. 1D).

**Fig. 3A** shows the Jablonski diagrams for the excited states of the +2 and 0 $O_CV_C$ and $V_CO_CV_C$ defects for their two dominant spin multiplicities: $s$ = 0 (blue) and 1 (green). The diagrams (calculated using emb-NEVPT2) show notable excitations within the visible spectrum between 2 and 3 eV (620 – 410 nm). The ground state of the +2 defects is non-degenerate with $^1A_1$ irreducible representation (irrep). The $O_CV_C^0$ is a triplet $^3A_2$, whereas $V_CO_CV_C^0$ has (nearly) degenerate singlet ($1^1A_1$) and triplet ($1^3B_2$) ground states. **Table S1** summarizes the singlet-triplet energy splitting for all defects, including the singlet-quintet splitting for the $V_CO_CV_C$s. The ground-state spin multiplicities alone disqualify $O_CV_C^0$ ($s$=1) and $V_CO_CV_C^0$ ($s$=0,1) from being the ST1 defect ($s$=0), however, we include them further in the analysis for comparison. **Table S1** also shows that for $V_CO_CV_C^{2-}$, the singlet, triplet, and quintet are all nearly degenerate, hence, we did not pursue this charge state further.

Both $O_CV_C^{2+}$ and $O_CV_C^0$ have first vertical excitation energies (VEEs) greater than 2.6 eV, being excited to doubly degenerate first excited states $^1E$ ($3a_1^1 2e^1$ and $3a_1^0 2e^2$) and $^3E$ ($3a_1^1 2e^3$), respectively (**Fig. S2**). These transitions are predicted to be bright, with transition dipole moments ($\mu_t$) of magnitude around 5 D, corresponding to oscillator strengths (*f*) ~ 0.4 - 0.6 (sum of the *f*s of the $^3E$ states) and spontaneous emission lifetimes ($\tau$) ~ 4 - 6 ns, which are similar to $N_CV_C^-$ (**Table 1**). *f* refers to the transition probability with an ideal value of 1 for single electron excitations. Our predictions for $N_CV_C^-$ matches well the experiments (**Table 1**) providing credence to our method. The lowest-energy metastable spins for both charges have *E* symmetry and are below the first excited state of their respective ground state spin multiplicities. On the other hand, the first excited states of the metastable spins are above ($^3A_2$) and below ($^1A_1$) the first excited state of the ground state spin multiplicities. The excited state profile of $O_CV_C^0$ is qualitatively similar to $N_CV_C^-$, as expected.[9] Although VEEs are upper bound of the zero-phonon lines (ZPLs): VEEs do not account for the atomic relaxation following excitation and thus correspond to an electronically and vibrationally excited state; the first VEE of $O_CV_C^{2+}$ (2.83 eV) is much higher than ST1's ZPL (2.2 – 2.3 eV). The difference between VEE and ZPL, e.g., for $N_CV_C^-$ is only ~ 0.2 eV (**Table 1**).

**Table 1.** Properties of select excited state(s) of the ground-state spin multiplicity and zero-field splitting (ZFS) constants of relevant spin triplets of select defects.

| Property | $^a$exp. ST1 | $^b$exp. NV$^-$ | $^c N_CV_C^-$ | $^d O_CV_C^{2+}$ | $^d O_CV_C^0$ | $^e V_CO_CV_C^{2+}$ |
|---|---|---|---|---|---|---|
| **gs** [irrep] | - | $^3A_2$ | $^3A_2$ | $^1A_1$ | $^3A_2$ | $1^1A_1$ |
| **es** [irrep] | - | $^3E$ | $^3E$ | $^1E$ | $^3E$ | $1^1B_2$, $2^1A_1$ |
| **VEE** **{ZPL}** [eV] | {2.2 - 2.3} | ~ 2.18 {1.95} | 2.26 | 2.83 | 2.65 | 2.38, 2.52 (2.29, 2.47) |
| $\|\mu_t\|$ [D] | - | - | 4.8 | 5.5 | 4.9 | 4.7, 6.1 (4.6, 6.0) |
| *f* | - | - | 0.40$^f$ | 0.64$^f$ | 0.48$^f$ | 0.20, 0.36 (0.18, 0.34) |
| $\tau$ [ns] | ~ 9 - 9.5 | ~ 8 - 12 | 9.5 | 3.7 | 5.6 | 8.6, 4.1 (10.1, 4.6) |
| **triplets** [irrep] | - | $^3A_2$, $^3E$ | $^3A_2$, $^3E$ | $^3E$ | $^3A_2$, $^3E$ | $1^3B_1$, $1^3A_2$, $1^3A_1$, $1^3B_2$ |
| $^g\boldsymbol{D_{ZFS}}$ [GHz] | 1.13 | 2.9, 1.4 | 3.0, 1.4 | 1.3 | 3.0, 1.5 | 2.6, 2.6, 1.3, 1.3 |
| $^g\boldsymbol{E_{ZFS}}$ [GHz] | 0.14 | 0, 0.07 | 0.0, 0.0 | 0.0 | 0.0, 0.0 | 0.9, 0.9, 0.1, 0.1 |

$^a$refs. [4-7]. $^b$refs. [19-22]. $^c$ref. [10]. $^d$three-state averaged calculations. $^e$eight- and six-state averaged calculations, with the later in parentheses. **Table S2** compares the result from the two calculations. $^f$2 × due to double degeneracy of the excited state. $^g$calculated using single-state (spin-restricted) CASSCF spin-density matrix including only the dipolar spin-spin interactions[23,24] (negligible spin-orbit coupling contributions) using bare $C_{15}DF_{12}O_{12}$ and $C_{26}OF_{18}O_{15}$ clusters for $D_CV_C$ and $V_CO_CV_C$, respectively. CASSCF spin-density matrix was averaged over two states for the doubly degenerate *E* states.

The two lowest excited states of $V_CO_CV_C^{2+}$ ($s$ = 0) have VEEs of 2.38 and 2.52 eV corresponding to transitions from $1^1A_1$ to $1^1B_2$ and $2^1A_1$ from an eight-state averaged simulation, respectively, closely matching ST1's ZPL. A six-state averaged calculation (more accurate than the eight-state because of fewer states averaged but reveals less spectral features – see **Methods**) yielded 2.29 and 2.47 eV for the same excitations (**Tables 1** and **S2**). $1^1B_2$ and $2^1A_1$ have large $\mu_t$ moments that are parallel to [110] and [001] axes, respectively (**Fig. S3**). $2^1A_1$ is shorter-lived and brighter than $1^1B_2$ ($\tau$ = 4 vs. 9 ns, $f$ = 0.36 vs. 0.20), and both are bright like $^3E$ of $N_CV_C^-$ (**Table 1**). **Fig. 3B** shows the dominant electron configurations (in the basis of the canonical orbitals in **Fig. 1D**) for the ground and four excited singlet states. Consistent with Brillouin's theorem, correlations for the ground state $1^1A_1$ are predominantly double excitations (**Figs. 3B, S4**). $1^1B_2$ and $2^1A_1$ correspond to the one-electron excitations from the $3a_12b_2$ to $4a_13b_2$ orbitals, whereas the higher energy $1^1A_2$ (dark) and $1^1B_1$ (bright) states are from $3a_12b_2$ to $2b_11a_2$ orbitals (**Fig. 3B, S4**). States that are higher ($3^1A_1$ and beyond) primarily involve double excitations from $3a_12b_2$ (**Fig. S4**).

There are four nearly degenerate triplet states below $1^1B_2$. The lowest two triplets are linear combinations of $1^3B_1$ and $1^3A_2$, whereas the higher two are combinations of $1^3A_1$ and $1^3B_2$. Both $1^3A_1$ and $1^3B_2$ are symmetric about $\sigma_v(1\bar{1}0)$ reflection plane (**Fig. 1B**), their combinations are thus labelled $1^3A'$ and $2^3A'$ (**Fig. 3A**), whereas $1^3B_1$ and $1^3A_2$ are anti-symmetric about $\sigma_v(1\bar{1}0)$, their combinations are thus labelled as $1^3A''$ (lowest energy triplet) and $2^3A''$ (**Fig. 3A**). The ground-state atomic structure appears to be an avoided crossing for the low-lying triplets causing the symmetry-allowed degenerate states to mix and split, albeit with small splitting energies of 0.06 ($1^3A'$ vs. $2^3A'$) and 0.02 eV ($1^3A''$ vs. $2^3A''$). Although state degeneracy is not expected for $C_{2v}$ (no $E$ or $T$ representation), the interaction between the orbitals of the two $V_C$s (e.g., the symmetric vs. anti-symmetric interactions about $\sigma_v(110)$) that ought to break state degeneracies appears to be very weak, leading to incidental degeneracies. $1^3A_1$ and $1^3B_2$ are spin-flip configurations of $2^1A_1$ and $1^1B_2$ (**Fig. 3B**), whereas $1^3A_2$ and $1^3B_1$ are spin-flip configurations of $1^1A_2$ and $1^1B_1$ (**Fig. S5**). Finally, $2^3B_2$ is dominated by configurations that populates the $2b_11a_2$ with two electrons (**Fig. S5**).

The spin singlet and triplet spectra of $V_CO_CV_C^0$ almost mirror each other within a ~ 2 eV energy window and are starkly different from $V_CO_CV_C^{2+}$ (**Fig. 3A**). However, the triplet spectrum has four more states than the singlet at ~ 1 eV: nearly degenerate $1^3A'$ and $2^3A'$ (mixture of $3^3B_2$ and $2^3A_1$) at 0.99 eV, and degenerate $2^3B_1$ and $2^3A_2$ at 1.2 eV (**Figs. S6**). The ~2 eV transition are the 4th ($3^1A_1$) and 9th ($4^3B_2$) excited states for the singlet and triplet, respectively (**Fig. 3A**). There are three low energy excited states: 0.45 eV ($2^1A_1$ and $2^3B_2$) and below (**Figs. S6**). All excited states are dark for both spin multiplicities with $f \leq 10^{-5}$, where $3^1A_1$ and $4^3B_2$ have $f \sim 10^{-7}$. These low $f$s indicate $V_CO_CV_C^0$ to have no measurable absorption and emission at and within 2 eV. However, it also means $V_CO_CV_C^0$ won't interfere with the optical detection of $V_CO_CV_C^{2+}$.

Finally, from optically detected magnetic resonance (ODMR) spectra, Lee et al. measured ZFS parameters of $D_{ZFS}$ = 1.135 GHz and $E_{ZFS}$ = 0.139 GHz for the triplet ST1.[4] A non-zero $E$ indicates lack of axial symmetry. We calculated the ZFS parameters for the relevant spin triplets of select defects (**Table 1**) resulting from the dipolar spin-spin interaction[23] (weak spin-orbit coupling). We use instead the bare capped clusters, which allowed for cluster reorientation and thus enable imposition of symmetry and specification of irreps. Our calculated ZFS constants for $N_CV_C^-$ match well the experimental values for both the $^3A_2$ and

$^3E$, and are similar to $O_CV_C^0$. ST1 having $E_{ZFS} \neq 0$ again disqualifies $O_CV_C^{2+}$ (due to its $C_{3v}$ symmetry) but provide further support for $V_CO_CV_C^{2+}$ to be the ST1 defect. Specifically, the spin triplets $1^3B_2$ and $1^3A_1$ are spin-orbit coupled with bright states $1^1B_2$, $2^1A_1$, and $1^1B_1$ (proportional to the radiative decay rate of the triplets to $1^1A_1$, **SI Note1, Table S3**) and both have calculated ZFS constants closely matching ST1's (**Table 1**). $1^3B_2$ and $1^3A_1$ have principal ZFS axes ($z$) along the $[\bar{1}10]$ (**Table S4**). $1^3B_2$ and $1^3A_1$ are consistent with the ODMR aligned axial magnetic field of 35(1)° relative [111], i.e., ~ [110], $[1\bar{1}0]$, $[\bar{1}10]$, $[\bar{1}\bar{1}0]$, for ST1,[4] further suggesting $1^3B_2$ and $1^3A_1$ to be most likely contributors in the observed triplet phosphorescence.

*Favorable coupling of $O_C$ with $V_C$s explains emergence of higher order defects.* **Fig. 4A** shows the calculated $O_C$ -$V_C$ coupling energies, which are very negative for both the formation of $O_CV_C$ (-5.13 eV) and $V_CO_CV_C$ (-4.28 eV), such that even at ppb-level $V_C$ and $O_C$ concentrations, the formation of $V_CO_CV_C$ should remain spontaneous. **Fig. 4A** also shows that $N_C$-$V_C$ coupling to $N_CV_C$ is slightly less negative compared to $O_CV_C$. However, coupling of $N_CV_C$ with another $V_C$ to $V_CN_CV_C$ is dramatically diminished (-4.60 to -0.76 eV). $V_CN_CV_C$ is unstable toward the formation of $N_CV_CV_C$ (-0.76 vs. -3.79 eV), conversely, $V_CO_CV_C$'s conversion to $O_CV_CV_C$ is unfavorable (-4.28 vs. -3.75 eV). The favorable $O_C$-$V_C$ coupling holds true for $O_C$'s ionized states: +1 and +2 (-4.83 and -4.76 eV), unlike $N_C^+$, which is greatly reduced (-2.01 eV). $V_CO_CV_C^{+/2+}$ are also predicted to form spontaneously from $O_CV_C^{+/2+}$ (-3.51 and -3.60 eV).

For $O_C$, $O_CV_C$, and $V_CO_CV_C$ to ionize, they must lose electrons to the holes in the valence band of a hole doped diamond or transfer electrons to electron acceptor co-defects/dopants following irradiation. **Fig. 4B** summarizes the calculated ionization energies (IE) minus the electron affinity of diamond (EA = $-\varepsilon_{\mathrm{CBM}}$), i.e., electron is promoted to the conduction band minimum (CBM) or (IE+$\varepsilon_{\mathrm{CBM}}$), for 0 and + of $O_C$, $O_CV_C$, and $V_CO_CV_C$. **Fig. 4B** also presents (IE+$\varepsilon_{\mathrm{CBM}}$) for $N_C^0$ as well as $N_CV_C^0$ and $V_C^0$ for comparison. $N_C^0$ is known (and predicted) to be a donor (exp. 1.62,[25] calc. 1.81 eV), whereas $O_C^0$ (one more valence electron than $N_C^0$) and $O_C^+$ (isoelectronic to $N_C^0$) are harder to ionize (3.17 and 4.78 eV). Their ionization involves removal of an electron from O/N-C anti-bonding $\sigma$ orbital. However, it is easier to ionize $O_CV_C$ than $N_CV_C$ and $V_C$ (3.47 vs. 4.40 and 4.48 eV) – an electron removed from a C[3c]-derived non-bonding orbital. And even though $V_CO_CV_C$ is harder to ionize to +1 than $O_CV_C$: 4.25 vs. 3.47 eV (**Fig. 4B**), the second ionization of $V_CO_CV_C$ is only slightly harder than the first and slightly easier than that for $O_CV_C^+$ (4.75 vs. 4.84 eV). $O_CV_C^+$ and $V_CO_CV_C^+$ are only slightly harder to ionize to +2 than $V_C^0$ to +1, therefore, if $V_C^+$ can exist, then both $O_CV_C^{2+}$ and $V_CO_CV_C^{2+}$ may exist.

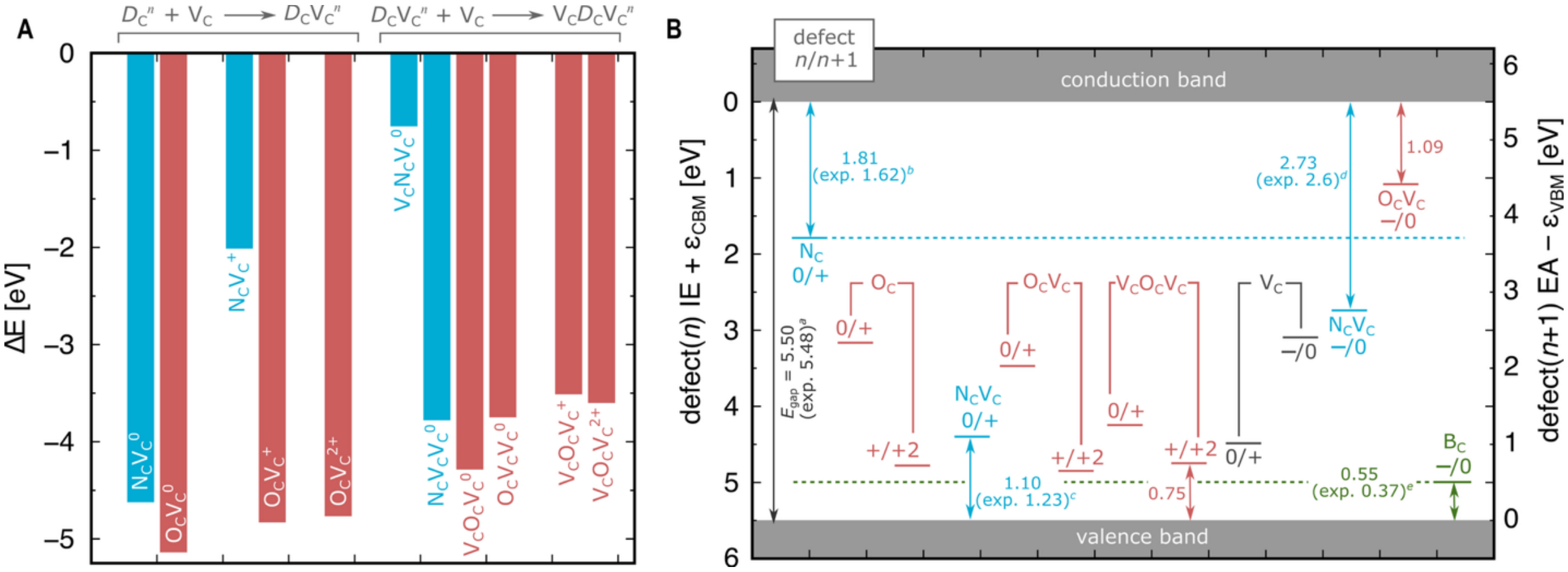

**Fig. 4. Coupling and ionization energies. A**, dopant-vacancy coupling energy for substitutional atom $D_C$ ($D$ = N and O) of charge $n$ with up to two $V_C$s. **B**, defect ionization energies (IE) minus electron affinity of diamond (EA = $-\varepsilon_{\text{CBM}}$, $\varepsilon_{\text{CBM}}$ is the energy of the conduction band minimum) or (IE+$\varepsilon_{\text{CBM}}$) for select defects (as labelled) transitioning from $n \rightarrow n$+1 charge state. The defect electron affinity (EA) plus ionization potential of diamond (IP = $-\varepsilon_{\text{VBM}}$, $\varepsilon_{\text{VBM}}$ is the energy of the valence band maximum) or (EA$-\varepsilon_{\text{VBM}}$) for $n$+1 → $n$ is simply $E_{\text{gap}}$–(IE+$\varepsilon_{\text{CBM}}$), where $E_{\text{gap}}$ is diamond's (indirect) band gap. (IE+$\varepsilon_{\text{CBM}}$)s and (EA$-\varepsilon_{\text{VBM}}$)s for some defects are annotated in eV. Experimental data from [a]ref. [26], [b]ref. [25],[c]ref. [27], [d]ref. [28], and [e]ref. [29]. In both **A** and **B**, energies were calculated within DFT-HSE06 with Coulomb corrections using DFT-r$^2$SCAN-L-optimized defective periodic 512-atom diamond supercells. The calculated $E_{\text{gap}}$ is from a quasi-particle $G_0W_0$ calculation for an 8-atom pristine diamond cell from ref. [10].

**Fig. 4B** also shows (IE+$\varepsilon_{\text{CBM}}$) for $N_CV_C^-$ and $O_CV_C^-$, as well as $V_C^-$ and $B_C^-$, which are common defects in diamond.[30] $V_C^0$ and $N_CV_C^0$ are viable electron acceptors/traps and ineffectual electron donors. Presence of $VC^0$s and $N_CV_C^0$s therefore will stabilize ionization of $O_CV_C$ and $V_CO_CV_C$ by acting as electron traps. $O_CV_C^0$ is a very deep electron acceptor that produces a shallow donor $O_CV_C^-$ (1.09 eV, **Fig. 4B**) . $O_CV_C^-$ is isoelectronic to $N_CV_C^{2-}$, which is a charge state that has never been observed. $B_C^0$ is a shallow electron acceptor and can be charged to $B_C^-$ from thermal excitation of electrons from the valence band maximum (VBM),[29,31] $O_CV_C$ and $V_CO_CV_C$ can then capture these free holes and ionize. Considering thermodynamics alone, defect states of charge $n$+1 below $N_C^{0/+}$ may act as electron acceptor for $N_C^0$ (**Fig. 4B**) and clearly detrimental to the ionization of many defects. Conversely, defect states of charge $n$ above $B_C^{-/0}$ may act as electron donor for $B_C^0$ and thus favor ionization of many other defects including $O_CV_C^0$ and $V_CO_CV_C^0$ all the way to +2.

**Discussion.** To our knowledge none of the structures put forth heretofore by experimentalists[4-7] and theoreticians[8,9] include $V_CO_CV_C$ as possible structure of ST1, and despite these efforts none have definitively identified the structure and charge state of ST1. Previously proposed structures include $O_C$,[9] $O_CV_C$,[9] $O_CC_xV_C$ ($x$ = 1-3),[5,8] $O_C(H_i)_x$ ($x$ = 1-2),[9] $O_CV_C(H_i)_x$ ($x$ = 1-3),[9] $O_CCV_CV_C$,[7] $V_CCO_CCV_C$,[7] and $O_CV_CO_C$[6] – all structures place $O_C$ in a three- or four-fold coordinated environment and none of the ones that has been looked at computationally could explain ST1's properties. Results from the simulations support $V_CO_CV_C^{2+}$ to be likely ST1. The predicted optical and magnetic properties of $V_CO_CV_C^{2+}$ explain the observed properties of the ST1 that no other O-based defect model has been able to heretofore. The two adjacent vacancies are promoted by the two lone pairs of $O_C$ in $V_CO_CV_C$, the same way the lone pair of a three-fold covalently coordinated $N_C$ promotes an adjacent vacancy in $N_CV_C$. The very thermodynamically favorable formation of $V_CO_CV_C$ relative to the isolated $O_C$, $V_C$, and $O_CV_C$, can significantly reduce free $O_C$ and $O_CV_C$ during vacuum annealing at high temperatures due to $V_C$ migration and eventual capture by $O_C$, especially at relatively low $O_C$ content and initially high $V_C$ concentrations: $[O_C] < [V_C]_i/2$. However, because the second $O_C$-$V_C$ coupling energy is less negative than the first, if $[O_C] \geq [V_C]_i$, then $O_CV_C$'s may emerge. The lack of bright optical signal for $O_CV_C^0$ may be due to its higher absorption energy or simply its absence due to dominance of the optically "dark" $V_CO_CV_C^0$. Irradiation of the O-containing samples that are hole-doped or with defect electron traps, e.g., samples with natural high B content or co-implanted with B, may be crucial in ionizing $V_CO_CV_C$ to +2. Upward surface band bending that causes hole accumulation can also potentially support ionization of near-surface $V_CO_CV_C$s.[31,32] Lastly, it is also important that electron donor impurities, e.g., $N_C$s, are absent for the +2 state to be stabilized.

**Data Availability**

All raw data used to generate figures and tables that appear in main text and SI are available upon reasonable request to be sent to the corresponding author.

**Code Availability**

VASP v. 6.4, Molpro v. 2024.1.1, and ORCA 6.1.0 used in this work are commercial codes that require user licenses. VASP subroutines with embedding implementation and associated Python scripts, and the standalone embedding integral generator code used to transform the embedding potential from Cartesian grid to GTO bases, are publicly available via GitHub: https://github.com/EACcodes/VASPEmbedding and https://github.com/EACcodes/EmbeddingIntegralGenerator, respectively, both under the Mozilla Public License 2.0.

**Acknowledgements**

The research described in this paper was conducted under the Laboratory Directed Research and Development (LDRD) Program at Princeton Plasma Physics Laboratory, a national laboratory operated by Princeton University for the U.S. Department of Energy under Prime Contract No. DE-AC02-09CH11466. The simulations presented in this article were performed on Princeton University's Tiger HPC and Princeton

Plasma Physics Laboratory's Stellar HPC resources managed and supported by Princeton Research Computing, a consortium of groups including the Princeton Institute for Computational Science and Engineering (PICSciE) and the Office of Information Technology's High Performance Computing Center and Visualization Laboratory at Princeton University.

**Author Contributions**

JMPM conceptualized, generated and interpreted the data, and wrote the manuscript.

**Competing interests**

The author declares no financial or non-financial competing interests

**Methods**

*Periodic DFT.* We used the Vienna Ab-Initio Simulation Package (VASP) v. 6.4[33,34] to perform periodic spin-restricted/unrestricted DFT within the projector augmented-wave (PAW) method and frozen core approximation. In conjunction, we used planewave basis set (kinetic energy cut-off of 660 eV) and standard VASP PAW potentials for all elements with frozen core 1s orbital.[35] We used the Gaussian smearing method with width 0.01 or 0.001 eV. We used smaller smearing width for defects with near degenerate in-gap states where larger smearing can introduce superfluous metallization. We introduced defects into pure 128-, 250-, and 512-atom periodic diamond supercells (**Fig. S7**). We used the 128- and 250-atom supercells to model the $O_CV_C$ and $V_CO_CV_C$ defects, respectively, that were subsequently used for capped-DFET simulations (see *Embedding potential*). We concurrently used Γ-point-centered 3×3×3 *k*-point sampling meshes for the two supercells. We used the largest 512-atom supercell to calculate the dopant-vacancy coupling and ionization energies (see *Dopant-vacancy coupling energy and (IE+$\varepsilon_{CBM}$)*), which is very sensitive to system size. Accordingly, we sampled only the Γ-point in the reciprocal space. The structures were fully optimized (atomic positions and cell vectors) without symmetry except for time-reversal within DFT-r$^2$-SCAN-L[15] with very tight absolute atomic force threshold of 0.001 eV/Å (because the cell vectors are being optimized) for the pure and defective 128- and 250-atom supercells and pure 512-atom supercell (**Table S5**). For the defective 512-atom supercells, we use the lattice vectors of the optimized pure supercell and subsequently relaxed only the atomic positions upon introduction of the defects with an absolute atomic force threshold of 0.01 eV/Å. For defects with even electrons, we performed optimization for two or three spin states where we constrained the system to have 0 (open- and closed-shell singlet), 2 (triplet), or in some cases 4 (quintet) unpaired electrons, but used only the structure of the lowest-energy spin in subsequent calculations. For odd number of electrons, we performed only the 1 unpaired electron (doublet) case (given the defects break $T_d$ symmetry and thus would not exhibit triple orbital degeneracies). We used single-point self-consistent DFT-HSE06[16] to obtain the final reference electron density for the embedding optimization (see *Embedding potential*) and total energies for the reaction energies (see *Dopant-vacancy coupling energy and (IE+$\varepsilon_{CBM}$)*). For all DFT-HSE06 calculations, the exact-exchange kernel was evaluated only at the Γ-point.

*Embedding potential.* We optimized the embedding potential ($V_{emb}(r)$) following the capped density functional embedding theory (capped-DFET) scheme,[11] which is an extension of the original DFET.[13] We first generated $C_{15}OF_{12}O_{12}$ and $C_{26}OF_{18}O_{15}$ clusters to model $O_CV_C$ and $V_CO_CV_C$, respectively (**Fig. 2**), by capping the defect clusters (*cl*) carved from bulk periodic models (vide infra). Specifically, $C_{15}O$ and $C_{26}O$ from periodic $C_{126}O$ and $C_{247}O$ supercells, respectively, with F and O added for terminal (monovalent) and bridging (divalent) capping, respectively. For the remaining atoms in the bulk, i.e., the environment (*env*), we used B and O for trivalent and divalent capping. We then obtain an optimized $V_{emb}(r)$ by maximizing the extended Wu-Yang functional[36] ($W$) using the full-system DFT-HSE06 electron density ($\rho^{full}$) as reference

$$W = E_{DFT}^{cl+cap1}[V_{emb}] + E_{DFT}^{env+cap2}[V_{emb}] - \int V_{emb}(r)\,(\rho^{full} + \rho^{cap1+cap2})\,dr \qquad \textbf{Eq. 1}$$

where, $E_{DFT}^{x+capy}$ and $\rho^x$ are the self-consistent DFT energies of fragment $x$ capped with elements *y* subject to $V_{emb}$. $\rho^{cap1+cap2}$ is the self-consistent electron density of an auxiliary fragment composed of only the capping elements combined into one species. The valence complementarity of the capping elements yields a closed-shell auxiliary fragment. The one-electron Hamiltonian ($H^0$) of the capped *cl* and *env* are modified as following

$$H_{emb}^0 = H^0 + V_{emb} \qquad \textbf{Eq. 2}$$

**Eq. 1** is equivalent to solving for $V_{emb}(r)$ that would yield

$$\rho^{cl+cap1}(\boldsymbol{r}) + \rho^{env+cap2}(\boldsymbol{r}) - \rho^{cap1+cap2}(\boldsymbol{r}) = \rho^{full}(\boldsymbol{r}) \qquad \textbf{Eq. 3}$$

$V_{emb}(r)$ thus account for the effective potential due to the mutual interaction of the cluster and environment in the full system that the capping elements do not capture. We used a modified VASP v 5.4 code to perform the $V_{emb}(r)$ optimization.[37,38] We use $V_{emb}(r)$s optimized for each defect type and charge state (**Fig. S8**) in the subsequent CASSCF and NEVPT2 calculations with the capped clusters.

*VEEs from emb-CASSCF and emb-NEVPT2.* We performed CASSCF and NEVPT2 using Molpro v. 2024.1.1. We used atom-centered Gaussian-type orbital (GTO) basis sets with complementary auxiliary bases for the evaluation of the two-electron integrals (resolution of identity approximation). Specifically, we used augmented correlation-consistent polarizable valence double $\zeta$ (aug-cc-pvdz)[39] for C and $O_C$. For capping elements, O and F, in $O_CV_C$, we also used aug-cc-pvdz, whereas in $V_CO_CV_C$, to manage computational cost, we used a smaller vdz bases (**Table S6**). We express the VASP-optimized $V_{emb}$ from real-space Cartesian grid to the GTO bases

$$V_{emb}^{i,j} = |\varphi_i\rangle\langle\varphi_i|V_{emb}|\varphi_j\rangle\langle\varphi_j| \qquad \textbf{Eq. 4}$$

by evaluating $\langle\varphi_i|V_{emb}|\varphi_j\rangle$s, where $\varphi_i$, $\varphi_j$ are the GTO basis functions, via the embedding integral generator code (EmbeddingIntegralGenerator).[40] We then add $V_{emb}^{i,j}$ to the one-electron Hamiltonian (**Eq. 2**). From a prior study, we found that use of a triple $\zeta$ basis (aug-cc-pvtz) leads to only ~ 0.01 eV change in the calculated VEEs of $N_CV_C^-$.[10] We performed $n_s$-state averaged CASSCF to include static correlation and to calculate for the ground and valence excited states without imposing symmetry. $n_s$ refers to the number of roots. We included the dynamic correlation via a multireference second-order perturbation theory method, namely, NEVPT2,[18] which was proven in the literature to be robust in determining excitation energies and relative energies of different spins with no semi-empirical input.[18,41] For $O_CV_C^{2+/0}$, $n_s$ equals 3 for spin-singlet and -triplet. For $V_CO_CV_C^{2+}$, $n_s$ equals 8 and 5 for spin-singlet and -triplet, respectively. For $V_CO_CV_C^0$, $n_s$ equals 5 and 9 for spin-singlet and -triplet, respectively. These $n_s$ values ensure that we include all the low-lying valence excited states or to reach at least a VEE of ~ 2 eV (as in $V_CO_CV_C^0$). On the other hand, the fewer states included in the state-averaged CASSCF, the better the (compromise) optimized orbitals are as basis for the wavefunction expansion of each state. However, for the spin-singlet $V_CO_CV_C^{2+}$, we found that we need at least $n_s = 8$ to include also the $1^1A_2$ and $1^1B_1$ states, which are the 6th and 7th excited states (7th and 8th roots) in emb-CASSCF, but becomes the 3rd and 4th excited states after adding dynamic correlation via emb-NEVPT2. We performed sensitivity test of VEE, $\mu_t$, $f$, and $\tau$ with respect to $n_s$ for $1^1A_1 \rightarrow 1^1B_2$, $2^1A_1$, and $3^1A_1$ of the spin-singlet $V_CO_CV_C^{2+}$. **Table S2** summarizes these quantities from $n_s = 6$ and 8 calculations. We report both the result from $n_s = 6$ and 8 in **Table 1**. For the spin-triplet $V_CO_CV_C^{2+}$, we needed $n_s = 5$ to include $2^3A''$. States $1^3A''$ and $2^3A''$ are respectively the 4th and 5th spin-triplet states (roots) in emb-CASSCF but becomes 3rd and 4th states in emb-NEVPT2.

We calculated the VEEs for the same spin multiplicity as the energy difference between the $V_{emb}$-embedded capped cluster energies of the $n$th excited ($E_{emb-NEVPT2}^{cl+cap1}(n)$) and ground electronic states ($E_{emb-NEVPT2}^{cl+cap1}(n=0)$) from emb-NEVPT2

$$VEE(0 \rightarrow n) = E_{emb-NEVPT2}^{cl+cap1}(n) - E_{emb-NEVPT2}^{cl+cap1}(n=0) \qquad \textbf{Eq. 5}$$

To reference the energy of the less stable spin multiplicity to the true ground state, we performed two-state averaged CASSCF (or three state, if one of the lowest-energy state of a given spin multiplicity is doubly degenerate) between the ground and higher energy spin multiplicity, followed by NEVPT2. Accordingly, the singlet-triplet (S-T) and singlet-quintet (S-Q) energy splitting are simply

$$\Delta E_{\mathrm{S-T\ or\ S-Q}} = E_{emb-NEVPT2}^{cl+cap1}(s = 1 \text{ or } 2) - E_{emb-NEVPT2}^{cl+cap1}(s = 0) \qquad \textbf{Eq. 6}$$

This energy difference (**Table S1**) is then added to the VEEs of the higher energy spin multiplicity to reference all their states to the true ground state.

In state-averaged (SA) CASSCF, each spin multiplicity was calculated using spin-pure configuration state functions (CSFs) in the many-body wavefunction expansion, except for when calculating for the S-T and S-Q splitting energy, and spin-orbit coupling (vide infra). We performed active space (AS) optimization using the molecular orbital theory as guide to determine the appropriate correlation orbital space. We include all C[3c]-derived non-bonding orbitals that engender the defects' electronic magnetization: there are three and six for $O_CV_C$ and $V_CO_CV_C$, respectively (gold box, **Fig. 1C,D**). These correspond to ASs of two or four electrons in three orbitals or (2e/4e,3o) for $O_CV_C^{2+/0}$ and (4e/6e/8e,6o) for $V_CO_CV_C^{2+/0/2-}$. For $O_CV_C^{0}$ and $V_CO_CV_C^{2-}$, we found it necessary to respectively include three and two additional correlating virtual orbitals in the AS, for a total of (4e,6o) and (8e,8o), respectively. We used CI coefficients ($c_i$) as a measure of orbital importance and using the following criterion: for any state, a virtual orbital is occupied in any configuration that has an absolute CI coefficient that is greater than or equal to 0.05 in state-averaged emb-CASSCF. We also performed AS size convergence test by increasing the number of both occupied and virtual orbitals in the ASs for $O_CV_C^{2+}$ ($s$=0), $O_CV_C^{0}$ ($s$=1), and $V_CO_CV_C^{2+}$ ($s$=0) to yield (8e,9o), (10e,12o), and (10e,12o), respectively (**Fig. S9**). **Table S7** summarizes the VEEs, $\mu_t$s, and $\tau$s for the different AS sizes. We found from these tests that our conclusions are robust despite of the quantitative variations in the predictions induced by the AS size.

*Symmetry determination and labelling of states*. We use the continuous symmetry measure (CSM) to determine the symmetry of $C_{15}OF_{12}O_{12}$ and $C_{26}OF_{18}O_{15}$ clusters for the two charge states. In CSM by Zabrodsky, Peleg, and Avnir,[42] given a symmetry group, $G$, the symmetry measure is $S'(G)$

$$S'(G) = \frac{1}{n_p}\sum_{i=1}^{n_p}\left\|P_i - \widehat{P_i}\right\|^2 \qquad \textbf{Eq. 7}$$

where, $n_p$ is the number of points in a structure, and $\left\|P_i - \widehat{P_i}\right\|$ is the displacement of point $P_i$ in the structure needed to reach the ideal point $\widehat{P_i}$ so that the structure belongs to the symmetry group $G$. The $P_i$ vectors origin are placed in the center of the structure and normalized by the maximum $||P_i||$. $\left\|P_i - \widehat{P_i}\right\|$ is a measure of the deviation of structure from a target shape. Therefore, the structure belongs to symmetry group $G$ that minimizes $S'(G)$, with an ideal value of $S'(G) = 0$. It follows that each symmetry operation $O_G$ belonging to group $G$ will have $S'(O_G) = 0$ and max $\left\|P_i - \widehat{P_i}\right\| = 0$. We used the CSM implementation in VASPKIT.[43] **Table S8** summarizes max $\left\|P_i - \widehat{P_i}\right\|$ and $S'(G)$ for each cluster and charge.

We determined the symmetry of the SA-CASSCF canonical orbitals by visual inspection and using character tables for $C_{3v}$, $C_{2v}$, and $C_s$ (**Table S9**). The AS pseudo-canonical orbitals (**Figs. 1D, S1**), which are the unitary transformation of the AS orbitals resulting from the diagonalization of the state-averaged effective Fock operator, conveniently exhibit the symmetry of their irreps, which is not always guaranteed. Note that for cases that this is not true, these orbitals can be used to generate symmetry-adapted linear combinations (SALCs) to match irreps. We use these pseudo-canonical orbitals without further transformations to assign state irreps (Mulliken labels) in conjunction with the dominant electron configurations (**Figs. S3-S5**) of each state (in the basis of said SA-CASSCF canonical orbitals) along with the symmetry product table (**Table S10**).

*Optical and magnetic properties.* The transition dipole moment vector ($\boldsymbol{\mu_t}$) and its magnitude is

$$\mu_t = |\boldsymbol{\mu_t}| = \left[\left|\mu_{t,x}\right|^2 + \left|\mu_{t,y}\right|^2 + \left|\mu_{t,z}\right|^2\right]^{\frac{1}{2}} \qquad \textbf{Eq. 8}$$

where $\mu_{t,[x,y,z]}$ are the Cartesian components of the transition dipole moment vector, calculated from a multi-state SA-CASSCF. The $\mu_{t,[x,y,z]}$s are origin invariant because all the states are guaranteed to be orthogonal in SA-CASSCF, even though for a charged defect the dipole moment is origin dependent.

The oscillator strength (*f*) for an $i \to j$ transition is

$$f(i \to j) = \frac{2}{3}\frac{m_e}{e^2\hbar^2}\frac{g_j}{g_i} VEE(i \to j)|\boldsymbol{\mu_t}|^2 \qquad \textbf{Eq. 9}$$

where $VEE$ is the energy for the $i \to j$ excitation in ha (calculated from emb-NEVPT2), $\boldsymbol{\mu_t}$ is in e·bohr (from the reference SA-CASSCF), and $f(i \to j)$ is dimensionless.[44] $m_e$ is the rest mass of an electron, $e$ is elementary charge, and $\hbar$ is Planck constant divided by $2\pi$. In atomic units (a.u.), the constants are simply equal to 1. $g_i$ and $g_j$ are the degeneracy of states *i* and *j*, respectively. We use the average values of $VEE(i \to j)$ and $f$ for $g_j$ = 2. Finally, we calculated the natural lifetime ($\tau$), i.e., due to a single relaxation event (spontaneous emission) of the excited states as the reciprocal of the Einstein coefficient for spontaneous emission ($A$) from state *j* to lower energy state *i*

$$\frac{1}{A(j \to i)} = \tau(j \to i) = 2.418884 \times 10^{-8} \frac{c^3}{2n_r VEE(i \to j)^2 f(i \to j)} \frac{g_j}{g_i} \qquad \textbf{Eq. 10}$$

where *c* is the speed of light in vacuum or the reciprocal of the fine-structure constant in a.u. = 137.03604, $n_r$ is the refractive index of the medium (diamond, ~ 2.42 at 520 - 600 nm), $VEE$ is in ha, and $2.418884 \times 10^{-8}$ is the conversion constant for time in a.u. to ns.[44] Note that the $g_i/g_j$ ratios simply cancel between **Eqs. 9** and **10**.

We also calculated the ZFS tensor originating from the spin-spin dipolar interactions using SA-CASSCF spin density matrix[23,24] for the bare capped clusters (i.e., without $V_{emb}$) as implemented in the ORCA[45,46] code. We solve for $D_{ZFS}$ and $E_{ZFS}$ defined from the ZFS Hamiltonian

$$\hat{H}_{\mathrm{ZFS}} = D_{xx}\hat{S}_x^2 + D_{yy}\hat{S}_y^2 + D_{zz}\hat{S}_z^2 = D_{\mathrm{ZFS}}\left[\hat{S}_z^2 - \frac{1}{3}\hat{S}^2\right] + E_{\mathrm{ZFS}}(\hat{S}_x^2 - \hat{S}_y^2) \qquad \textbf{Eq. 11}$$

where, $D_{\mathrm{ZFS}} = 3D_{zz}/2$ and $E_{\mathrm{ZFS}} = (D_{xx} - D_{yy})/2$, in the principal axis system. $(D_{xx}, D_{yy}, D_{zz})$ are the diagonal elements of the diagonalized ZFS tensor - here we included only the spin-spin component. $\hat{S}_x$, $\hat{S}_y$, and $\hat{S}_z$ are the spin operators and $\hat{S}$ is the total spin operator. We used the method by Sinnecker and Neese[23] implemented in ORCA 6.1.0. For $O_CV_C^{2+/0}$, we used single- and two-state CASSCF-single-particle density matrixes for non-degenerate (irrep *A*) and doubly degenerate states (irrep *E*). For the spin triplets of $V_CO_CV_C^{2+}$, the ZFS tensor were calculated using single-state CASSCF-single-particle density matrixes for a given irrep.

*Dopant-vacancy coupling energy and $(IE + \varepsilon_{CBM})$.* The dopant-vacancy coupling energy is simply the energy for the reaction

$$D_{\mathrm{C}}(\mathrm{V_C})_m{}^n + \mathrm{V_C} \rightarrow D_{\mathrm{C}}(\mathrm{V_C})_{m+1}{}^n + \mathrm{C(diamond)}$$

Here we explore $m$ = [0,1] and $D$ = [O,N]. $n$ is the charge of the defect, which does not change during the reaction. For completeness, $\mathrm{C(diamond)}$ appears on the right hand of the chemical equation because one of the C[n*D*] moves to occupy the $V_C$ upon coupling, becoming a pristine diamond C atom.

To calculate the (IE $+\varepsilon_{\mathrm{CBM}}$), we first calculated the charge transfer energy ($\Delta E_{\mathrm{CT}}$) between a defect of charge $n$ ($X^n$) and an isolated $V_C$, the latter being the as the electron acceptor

$$X^n + \mathrm{V_C} \rightarrow X^{n+1} + \mathrm{V_C^-}$$

The $V_C/V_C^-$ pair thus effectively acts as a reference. We then calculate the (IE$+\varepsilon_{\mathrm{CBM}}$) of $\mathrm{V_C^-}$

$$\mathrm{V_C^-} \rightarrow \mathrm{V_C} + e^-(\mathrm{CBM})$$

where, $e^-(\mathrm{CBM})$ is the energy of an electron in the CBM at the very dilute carrier density limit (negative electron affinity [EA] of the host), which we approximate as the energy of the CBM of $V_C$ ($\varepsilon_{\mathrm{CBM}}$). We calculated an $(\mathrm{IE} + \varepsilon_{\mathrm{CBM}})_{\mathrm{V_C^-/V_C}}$ of $E_{\mathrm{V_C}} + \varepsilon_{\mathrm{CBM}} - E_{\mathrm{V_C^-}}$ = 3.09 eV. Finally, we calculate the (IE$+\varepsilon_{\mathrm{CBM}}$) of $X^n$

$$X^n \rightarrow X^{n+1} + e^-(\mathrm{CBM})$$

using the expression

$$(\mathrm{IE} + \varepsilon_{\mathrm{CBM}})_{X^n/X^{n+1}} = \Delta E_{\mathrm{CT}}(X^n + \mathrm{V_C} \rightarrow X^{n+1} + \mathrm{V_C^-}) + (\mathrm{IE} + \varepsilon_{\mathrm{CBM}})_{\mathrm{V_C^-/V_C}} \qquad \textbf{Eq. 12}$$

Conversely, to calculate for electron affinity (EA) of $X^{n+1}$ relative to the valence band maximum or VBM (negative of the ionization potential [IP] of the host)

$$X^{n+1} + e^-(\mathrm{VBM}) \rightarrow X^n$$

one may use the $(\mathrm{IE} + \varepsilon_{\mathrm{CBM}})_{X^n/X^{n+1}}$ and the following relationship

$$(\mathrm{EA}-\varepsilon_{\mathrm{VBM}})_{X^{n+1}/X^{n}} = E_{gap} - (\mathrm{IE}+\varepsilon_{\mathrm{CBM}})_{X^{n}/X^{n+1}} \quad \textbf{Eq. 13}$$

where is $E_{gap}$ the calculated IP–EA band gap of pure diamond from $G_0W_0$ (5.5 eV).[10] **Table S11** summarizes $\Delta E_{\mathrm{CT}}(X^{n} + \mathrm{V_C} \rightarrow X^{n+1} + \mathrm{V_C^-})$s for the defects presented in **Fig. 4B**.

For all the energies, we generated defective structures from the pristine 512-atom diamond supercell (**Fig. S7**). For charged defects, following Makov and Payne's method,[47] we introduced Coulomb corrections (monopole-monopole and quadrupole-monopole) for errors resulting from the interaction of the charged defects with their periodic images. This correction scheme proved to approach the value from more sophisticated correction schemes at sufficiently large cells, e.g., 512-atom supercells for diamond defects.[48] The monopole-monopole correction is given by

$$E_{\mathrm{M-M}} = \frac{e^2 q^2 \alpha}{2L\varepsilon} \quad \textbf{Eq. 14}$$

$q$ is the excess charge in the cell, $e$ is electron charge, $L$ is the length of the side of the (cubic) cell, $\alpha$ is Madelung constant (2.8373 for simple cubic), and $\varepsilon$ is the material's dielectric constant ($4\pi\varepsilon_0 = 0.069446\ e^2\mathrm{eV}^{-1}\text{Å}^{-1}$; $\varepsilon = 4\pi\varepsilon_0\varepsilon_r$; $\varepsilon_r$ is the dielectric of the pure material). The correction for monopole-quadrupole (M-Q) interaction is

$$E_{\mathrm{M-Q}} = \frac{2\pi e q (Q_q - Q_{q=0})}{3L^3\varepsilon} \quad \textbf{Eq. 15}$$

where,

$$Q = \int \rho(\boldsymbol{r})\, |\boldsymbol{r} - \boldsymbol{r_0}|^2 d^3\boldsymbol{r} \quad \textbf{Eq. 16}$$

evaluated for $q$-charged defect and neutral ($q$ = 0) pristine diamond with the integrand within the volume of supercell. $\rho(\boldsymbol{r})$ is the electron density at point $\boldsymbol{r}$ and $\boldsymbol{r_0}$ is an arbitrary origin. Here, $(Q_q - Q_{q=0})$ approximates the charged defect induced quadrupole moment. The corrected total energy of a charged cubic supercell with side of length $L$ ($E(L \rightarrow \infty)$) subject to a neutralizing jellium is thus

$$E(L \rightarrow \infty) = E(L) + \frac{e^2 q^2 \alpha}{2L\varepsilon} + \frac{2\pi e q (Q_q - Q_{q=0})}{3L^3\varepsilon} \quad \textbf{Eq. 17}$$

For neutral defects, we included a correction due to dipole-dipole interaction

$$E_{\mathrm{D-D}} = \frac{2\pi}{3L^3\varepsilon}\left|\int \rho(\boldsymbol{r})\,(\boldsymbol{r} - \boldsymbol{r}_0)\, d^3\boldsymbol{r}\right|^2 = \frac{2\pi}{3L^3\varepsilon}|\mu|^2 \quad \textbf{Eq. 18}$$

thus

$$E(L \rightarrow \infty) = E(L) + \frac{2\pi}{3L^3\varepsilon}|\mu|^2 \quad \textbf{Eq. 19}$$

We set $\varepsilon_r = 5.69$, which is the calculated $\varepsilon_r(\omega = 0)$ for the frequency-dependent dielectric constant for a 512-atom pure diamond. We used the Green-Kubo formula implemented within VASP[49] using DFT-r$^2$-SCAN-L, Lorentzian broadening of 0.1, summing over 1024 occupied and 1024 unoccupied bands, and using the perturbation expansion after discretization method for the evaluation of the orbital derivatives. Finally, $L$ = 14.2710 Å (**Table S5**). $E_{\mathrm{M-M}}$ is a positive correction that is equal to 0.252 and 1.006 eV for monovalent and divalent defects, respectively. $E_{\mathrm{M-Q}}$ is a negative correction that is around -0.07 to -0.08 eV and -0.30 to -0.31 eV for monovalent and divalent defects, respectively. $E_{\mathrm{D-D}}$ corrections for the neutral defects are all $\geq 0$ but $< 0.01$ eV.